\documentclass[aps,pra,superscriptaddress,twocolumn,showpacs]{revtex4}
\usepackage{amsmath}
\usepackage{amssymb}
\usepackage{bm}
\usepackage{epsfig}
\usepackage{graphicx}
\usepackage{color}

\newcommand{\re}[1]{(\ref{#1})}
\newcommand{\wh}[1]{\widehat{#1}}

\begin{document}

\title{Collapse of spin-orbit coupled Bose-Einstein condensates}
\author{Sh. Mardonov}
\affiliation{Department of Physical Chemistry, The University of the Basque Country, 48080 Bilbao, Spain}
\affiliation{The Samarkand Agriculture Institute, 140103 Samarkand, Uzbekistan}
\affiliation{The Samarkand State University, 140104 Samarkand, Uzbekistan}
\author{E. Ya. Sherman}
\affiliation{Department of Physical Chemistry, The University of the Basque Country, 48080 Bilbao, Spain}
\affiliation{IKERBASQUE Basque Foundation for Science, Bilbao, Spain}
\author{J. G. Muga}
\affiliation{Department of Physical Chemistry, The University of the Basque Country, 48080 Bilbao, Spain}
\affiliation{Department of Physics, Shanghai University, 200444 Shanghai, Peoples Republic of China}
\author{Hong-Wei Wang}
\affiliation{Department of Physics, Shanghai University, 200444 Shanghai, Peoples Republic of China}
\author{Yue Ban}
\affiliation{Department of Electronic Information Material, Shanghai University, 200444, Shanghai, People's Republic of China}
\author{Xi Chen}
\affiliation{Department of Physics, Shanghai University, 200444 Shanghai, Peoples Republic of China}

\begin{abstract}
{A finite-size quasi two-dimensional Bose-Einstein condensate collapses if the attraction between atoms is sufficiently 
strong. Here we present a theory of collapse for condensates with the interatomic attraction and spin-orbit coupling. 
We consider two realizations of spin-orbit coupling: the axial Rashba coupling and 
balanced, effectively one-dimensional,  Rashba-Dresselhaus one. In both cases 
spin-dependent ``anomalous'' 
velocity, proportional to the spin-orbit coupling strength, plays a crucial role. For the Rashba coupling, this velocity
forms a centrifugal component in the density flux opposite to that arising due to the attraction between
particles and prevents the collapse at a sufficiently strong coupling. 
For the balanced Rashba-Dresselhaus coupling, the spin-dependent velocity can spatially split the initial state in one
dimension and form spin-projected wavepackets, reducing the total condensate density. 
Depending on the spin-orbit coupling strength, interatomic attraction, and the initial state, this splitting either prevents
the collapse or modifies the collapse process. These results show that the 
collapse can be controlled by a spin-orbit coupling, thus, extending the domain of existence of condensates of attracting
atoms.} 
\end{abstract}

\maketitle

\section{Introduction}

Understanding Bose-Einstein condensates (BEC) of interacting particles
is one of the most interesting problems in condensed matter physics \cite{dalfovo1999}. 
For uniform three-dimensional systems repulsion between the bosons depletes
the condensate, while attraction leads to the condensate instability seen as 
the appearance of Bogoliubov modes with imaginary frequencies. For the finite-size
condensates this instability can be seen in their collapse \cite{kagan, cornish, abdullaev, kagan2, stoof1}.
The collapse, where the size of the state goes to zero after a finite time,
strongly depends on the spatial dimension $D$ and is possible only in the $D=2$ 
and $D=3$ condensates. The physics of the collapse is related to the fundamental
problems of nonlinear optics and quantum mechanics \cite{book}, plasma instability \cite{berge}, and 
polaron formation \cite{rashba_review}.

The main features of the collapse of a free, not restricted by an external potential, condensate, 
are determined by the interplay of its positive quantum 
kinetic and negative attraction energies dependent on the characteristic size of the condensate $a$. 
The kinetic energy is proportional to $a^{-2}$ while the attraction contribution 
behaves as $-a^{-D}$. For $D=3$ the dependence of
the total energy on $a$ is non monotonic and the collapse with $a\rightarrow 0$ occurs
at any interaction strength since at small $a$ the attraction dominates \cite{ueda1998}. 
For $D=2$ the interaction and kinetic energies scale as $a^{-2}$ and the collapse occurs only at a strong
enough attraction.

The BEC physics becomes much richer with synthetic gauge fields \cite{spielman2009} 
and synthetic spin-orbit coupling (SOC) \cite{wang2010,spielman2011}. For the latter, 
optically produced atomic pseudospin $1/2$ is coupled to atomic momentum and
to a synthetic magnetic field.
The SOC can be produced in various forms, simulating the Rashba and the Dresselhaus 
symmetries [\onlinecite{zhai2012, spielman2013}] {known in solid state physics.} 
This coupling opens a venue to the appearance of new phases in a variety of ultracold
bosonic \cite{artem, osterloh, ruseckas, galitski, dalibard,
ho2011,stringari,anderson, zhang2012, zhang} and fermionic \cite{liu,wang2012,cheuk2012,iskin2012} ensembles. 
It is well appreciated that the SOC plays crucial role in BEC physics 
in uniform three-dimensional gases with interparticle repulsion \cite{Ozawa,Barnett}. For $D=2$ the phases 
of the BEC of repelling bosons trapped in a harmonic potential were found in Ref. [\onlinecite{Sinha}].

{One of the advantages of cold atomic gases is the fact that due to a very large 
particle wavelength compared with the atomic radius, the interatomic interaction can be accurately
described by a single parameter, the scattering length $a_{s},$ where positive (negative) 
$a_{s}$ corresponds to repulsion (attraction) 
between the atoms. The attraction can be achieved
by means of the Feshbach resonance \cite{Chin} in a certain range of the system parameters. 
Here we study joint effect of the interatomic attraction and spin-orbit on the spread and collapse of 
a quasi two-dimensional spin-orbit coupled BEC.} 

This paper is organized as follows. In Sec. II we 
show {by qualitative arguments, a variational approach, and direct numerical solution of the Gross-Pitaevskii equation,
that the effect of the anomalous spin-dependent velocity due to the spin-orbit coupling \cite{Adams} can either completely 
prohibit the collapse or strongly modify the collapse process. We study the condensate dynamics and analyze the conditions at which the  
collapse does not occur. Possible relations to experiment and conclusions will be given in Sec. III.}

\section{Collapse in the presence of spin-orbit coupling}

\subsection{General formulation: Hamiltonian and the collapse process}

We consider a pancake-shaped condensate of pseudospin $1/2$ particles described
by a two-component wave function $\Psi =\left[\psi ^{\uparrow }(\mathbf{r},t),\psi^{\downarrow }(\mathbf{r},t)\right]^{T},$ 
where ${\bf r} \equiv (x,y),$ normalized to the
total number of particles $N\gg 1.$  
In the presence of the spin-orbit coupling, the evolution of the
wavefunction is described by a system of coupled nonlinear partial differential equations in the 
Gross-Pitaevskii-Schr\"{o}dinger form
\begin{equation}
i\hbar \frac{\partial {\Psi }}{\partial t}=\left[-\frac{\hbar^{2}}{2M}\Delta
+\wh{H}_{\mathrm{so}}
+\frac{1}{2}\left({\mathbf B}\cdot\wh{\bm\sigma} \right)
-g_{2}\left\vert \Psi \right\vert ^{2}\right]{\Psi }.
\label{SHE}
\end{equation}%
{Here $M$ is the particle mass, $\wh{H}_{\mathrm{so}}$ is the SOC Hamiltonian, 
${\mathbf B}$ is the effective magnetic
field, and  $\wh{\bm\sigma} =\left( \wh{\sigma} _{x},\wh{\sigma} _{y}, \wh{\sigma} _{z}\right) $ is
the spin operator. The coupling constant in Eq. \re{SHE} is given by $g_{2}=-4\pi\hbar^{2}a_{s}/Ma_{z},$ which we assume 
for simplicity to be spin-independent, where 
$a_{z}$ is the condensate extension along the $z-$ axis, and $a_{s}$ is negative \cite{kagan,abdullaev,kagan2}.  
Below we consider two strongly different forms of $\wh{H}_{\mathrm{so}}$:
the Rashba coupling with the spectrum axially symmetric in the momentum space and 
the balanced, essentially, one-dimensional Rashba-Dresselhaus coupling.}

Without loss of generality, we consider an initial state prepared in a parabolic potential at zero temperature as:
\begin{equation}
{\Psi }({\bf r},t=0) \equiv A(0)\exp \left[-\frac{r^{2}}{2a^{2}(0)}\right]{\bm \psi}(0),
\label{initial}
\end{equation}
where ${\bm \psi}(0)$ is the initial spinor, $A(0)=\sqrt{N/\pi}/a(0),$ and $a(0)$ is the initial width. 
At $t=0,$ the confining potential is switched off \cite{dalfovo1999} and the spin-orbit coupling and the 
attraction between the atoms are switched on. The subsequent dynamics is, thus, a response of the system 
to the instantaneous change in the potential, interaction, and spin-orbit coupling. 

In what follows we use the units $\hbar \equiv M\equiv 1$ and 
the dimensionless interaction $\widetilde{g}_{2}\equiv -4\pi{a_{s}}/a_{z}.$  
The unit of length $\ell$ can be
chosen arbitrarily, and the corresponding unit of time is $\ell^{2}.$

{We address first} the collapse without spin-dependent effects. 
Here the energy of the system is
\begin{equation}
\displaystyle E= -\frac{1}{2}
\int\left[\Psi^{\dagger}\Delta\Psi+\widetilde{g}_{2}\left\vert\Psi\right\vert^{4}\right]dxdy,
\label{TE}
\end{equation}%
and the evolution can be described by a variational approach based on Gaussian ansatz \cite{abdullaev} 
\begin{eqnarray}
{\Psi }({\bf r},t) = A(t)\exp \left[-\frac{r^{2}}{2a_{\rm v}^{2}(t)}\left(1+ib_{\rm v}(t)\right)\right]{\bm \psi}(0),
\label{wfup}
\end{eqnarray}%
where the {variational parameters} $b_{\rm v}(t)$ and $a_{\rm v}(t)$  are the chirp and the packet width, respectively. 
The equation of motion for $a_{\rm v}$ becomes $\overset{..}{a}_{\rm v}=-{\Lambda }/{a_{\rm v}^{3}}$, 
where $\Lambda =\left(\widetilde{g}_{2}N-\lambda_{\rm v}\right)/2.$ 
The collapse occurs if $\widetilde{g}_{2}N$ exceeds the  variational threshold value 
$\lambda_{\rm v}=2\pi$ \cite{critical}. The solution of this equation is 
\begin{equation}
a_{\rm v}(t)=a(0)\sqrt{1-\frac{\Lambda t^{2}}{a^{4}(0)}}.
\label{at}
\end{equation}
The timescale of the evolution is the collapse time $T_{c}\equiv a^{2}(0)/\sqrt{\Lambda},$
and the characteristic collapse velocity is $v_{c}\equiv a(0)/T_{c}=\sqrt{\Lambda}/a(0).$

The key point in the understanding of the role of the spin-orbit coupling in the collapse process 
is the modified velocity
\begin{equation}
{\mathbf v}={\mathbf k}+{\bm\nabla}_{\mathbf k}\wh{H}_{\rm so},
\label{vel1}
\end{equation}
with ${\mathbf k}=-i\partial/\partial{\mathbf r}$, including the anomalous velocity \cite{Adams} term ${\bm\nabla}_{\mathbf k}\wh{H}_{\rm so}$ 
(here ${\bm\nabla}_{\mathbf k}\equiv \partial/\partial {\mathbf k}$) directly 
related to the particle spin. 
The evolution of the probability density $\rho ={\Psi }^{\dagger }{\Psi }$ 
is given by the continuity equation
\begin{equation}
\frac{\partial \rho }{\partial t}+{\bm\nabla}\cdot\mathbf{J}(\mathbf{r},t)=0,
\end{equation}%
with the components of the flux density
\begin{equation}
\mathbf{J}(\mathbf{r},t)=\frac{i}{2}\left[{\Psi }{\bm\nabla}{\Psi }^{\dagger }-{\Psi }^{\dagger }{\bm\nabla}{\Psi}\right]
 +{\Psi }^{\dagger }\left[{\bm\nabla}_{\mathbf k}\wh{H}_{\rm so}\right]{\Psi}.
\end{equation}%
The spin components of the condensate are given by expectation values%
\begin{equation}
\left\langle \wh{\sigma} _{i}(t)\right\rangle =\frac{1}{N}\int {\Psi }^{\dagger }\wh{\sigma}
_{i}{\Psi dxdy}.
\end{equation}

\begin{figure}[t]
\includegraphics[width=80mm]{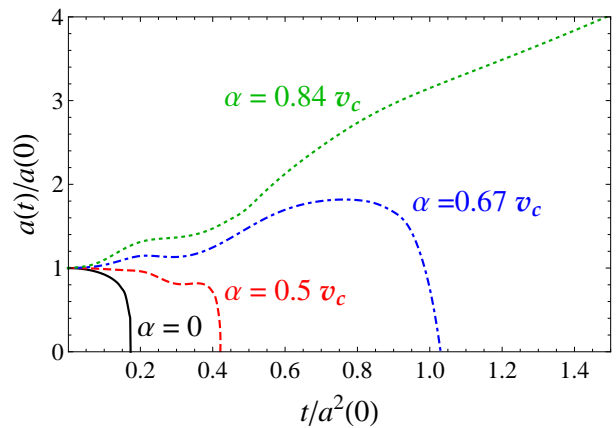} 
\caption{(Color online) Time dependence of the condensate width for
$\widetilde{g}_{2}N=16\pi$ and the values of $\alpha$
marked near the lines. The green short-dashed line corresponds to the absence of collapse.}
\label{figure1}
\end{figure}

\subsection{Rashba spin-orbit coupling} 

The first form of spin-orbit interaction that we consider is
the Rashba coupling
\begin{equation}
\wh{H}_{\mathrm{so}}\equiv\wh{H}_{R}=\alpha \left( k_{x}\widehat{\sigma }_{y}-k_{y}\widehat{\sigma }_{x}\right),
\label{rashba}
\end{equation}
{with the coupling constant $\alpha$ and ${\mathbf k}\equiv\left(k_{x},k_{y}\right).$} 
The corresponding spin-dependent terms in the velocity operators in Eq. \re{vel1} become
\begin{equation}
\frac{\partial \wh{H}_{R}}{\partial k_{x}}={\alpha}\widehat{\sigma }_{y} ,\quad
\frac{\partial \wh{H}_{R}}{\partial k_{y}}=-{\alpha }\widehat{\sigma }_{x}.
\label{velocity}
\end{equation}%
The spatial scale of the SOC  effects is described by the
characteristic distance the particle has to move to flip the spin,
$L_{\mathrm{so}}=1/\alpha$. 
The corresponding spin rotation angle at the particle displacement $L$ is of the order of $L/L_{\mathrm{so}}.$
At the initial stage of the BEC evolution $t\ll T_{c}$ we obtain from Eq. \re{SHE} for $\Psi({\bf r},t=0)$ in Eq. \re{initial} with ${\bm\psi}(0)=[1,0]^{T},$  
\begin{equation}
\frac{\partial }{\partial t}\psi ^{\downarrow }({\bf r},t\rightarrow 0)=i
\frac{\sqrt{N}}{\sqrt{\pi }a^{3}(0)}\frac{x+iy}{L_{\mathrm{so}}} 
\exp \left[ -\frac{{r^{2}}}{2a^{2}(0)}\right].
\end{equation}%
As a result, the $\psi ^{\downarrow }({\bf r},t)$ component begins to grow at 
distances $r\sim a(0)$ with a rate proportional to $\alpha.$ At a sufficiently large $\alpha$ 
this growth can eventually lead to the collapse prevention.

\begin{figure}[t]
\begin{center}
\includegraphics[width=80mm]{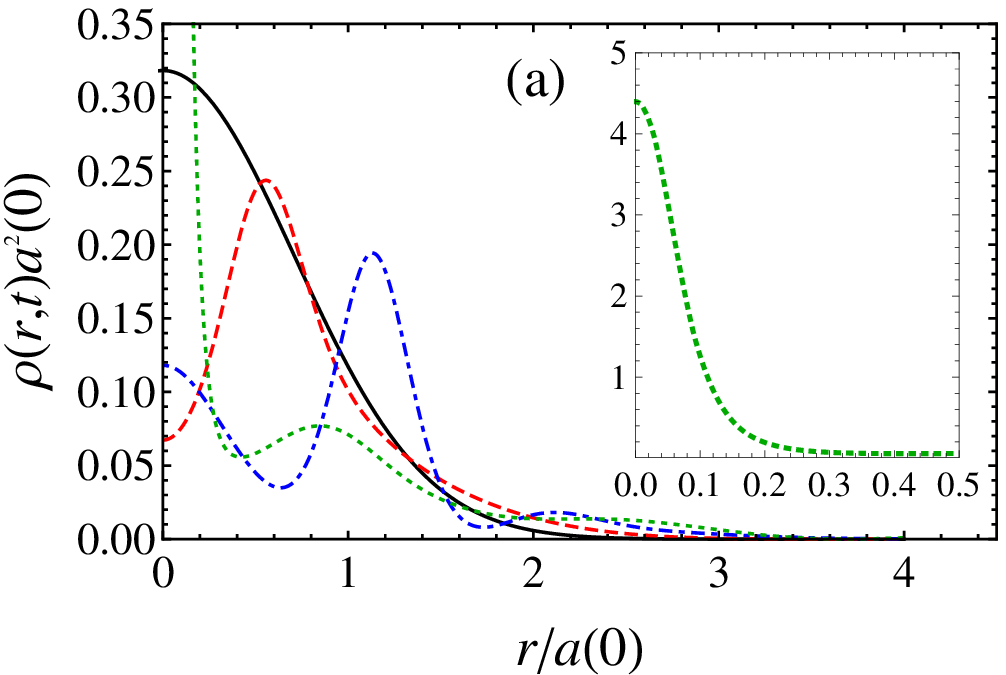}\\
\includegraphics[width=80mm]{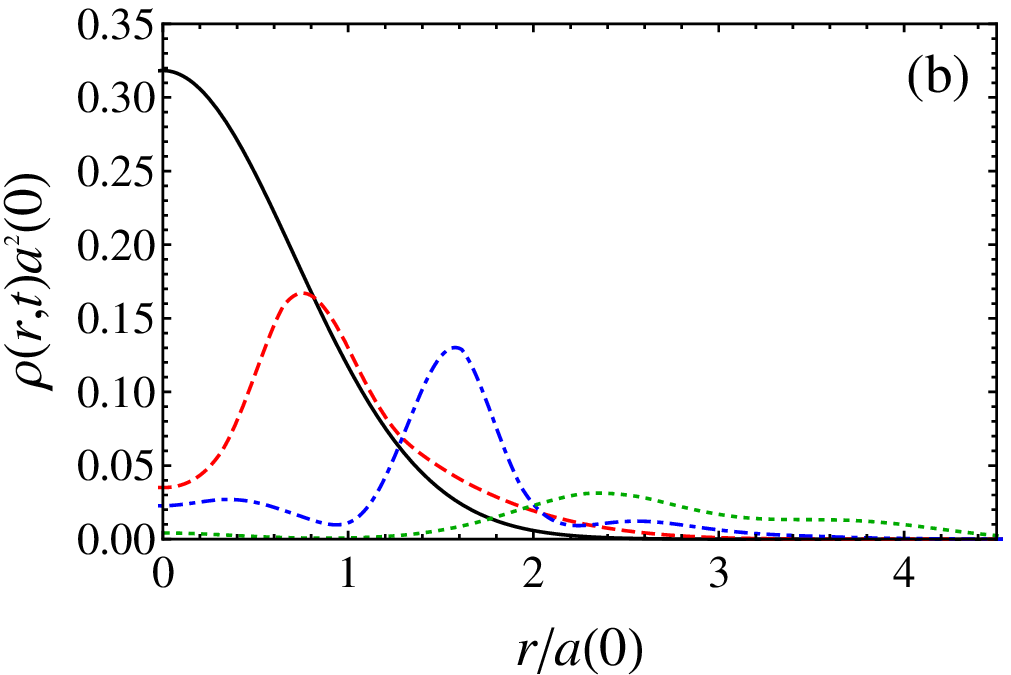}
\end{center}
\caption{ (Color online) Density profile $\protect\rho(r,t)$ for
$\widetilde{g}_{2}N=16\pi.$ Black solid
line is for $t=0$, red dashed line is for $t=0.2a^{2}(0)$, blue dot-dashed line
is for $t=0.4a^{2}(0)$, and green dotted line is for $t=a^{2}(0)$. 
(a) Here $\alpha=0.67 v_{c},$ and the dotted green line
shown in detail in the inset, clearly demonstrates the collapse.
(b) Here $\alpha =0.84 v_{c},$ the condensate is robust against attraction and can spread without collapsing.}
\label{figure2}
\end{figure}

At $t>0,$ the spatially nonuniform spin evolution begins. Since the spin precession angle 
at the displacement of $a(0)$ is of the order of $a(0)/L_{\rm so},$ starting 
from the fully polarized ${\bm\psi}(0)=\left[1,0\right]^{T}$ state, 
the atoms acquire the anomalous velocity of the order of 
$a(0)/L_{\mathrm{so}}\times \alpha\sim\alpha^{2}a(0)$ for the weak SOC $a(0)\ll L_{\mathrm{so}},$ or 
of the order of $\alpha$ otherwise.
The criterion of a large spin rotation in the collapse is $a(0)>L_{\mathrm{so}},$ 
that is $\alpha>1/a(0),$ while the condition of a sufficiently large developed 
anomalous velocity is $\alpha>v_{c},$ that is $\alpha>\sqrt{\Lambda}/a(0).$
If the latter inequality is satisfied, the centrifugal component in the flow caused by the SOC \cite{Cooper}
can prevent the collapse, as we explain in detail below. 
The condition of a weak effect of magnetic field on the 
collapse can be formulated {as smallness of spin precession angle due to the Zeeman splitting
compared to the precession angle due to the spin-orbit coupling, that is $T_{c}B\ll\min\left\{a(0)/L_{\mathrm{so}},1\right\}.$
We will assume this condition and neglect the effects of the Zeeman splitting.}

We begin the analysis of the joint effect of the SOC and the interatomic 
attraction with numerical results obtained by direct integration of Eq.\re{SHE}  
for a strong attraction, $\widetilde{g}_{2}N\gg 1,$ where the effect is clearly seen, 
taking the initial spin state
${\bm \psi}(0)=[1,0]^{T}$. Figure \ref{figure1} shows the time-dependent width of the packet defined as
\begin{equation}
a(t)\equiv \frac{N}{\sqrt{2\pi }}
\left[ \int \left\vert \Psi \right\vert ^{4}dxdy%
\right]^{-1/2},
\label{atime}
\end{equation}
where $\Psi$ is obtained by a direct solution of Eq. \re{SHE} for 
several values of $\alpha.$ The solid line in Fig. \ref{figure1} corresponds 
to the collapse at $\alpha=0$ where in the vicinity of $T_{c},$ the numerically calculated using Eqs. \re{SHE} and 
\re{atime}, width $a(t)$ is accurately described by variational Eq. \re{at} with $a(t)\sim(T_{c}-t)^{1/2}$.  

When spin-orbit coupling is included, the following features may be seen. (i) At short time $t\ll T_{c}$, 
the attraction-induced velocity develops linearly with $t$, while the
anomalous velocity increases as $t^{2}$. As a result, the $a(t)-$%
dependences for all values of $\alpha$ are the same at small $t$. 
(ii) {The packet width $a(t)$ increases with time}, reaches a plateau, and then
decreases to zero. Thus, with the increase in $\alpha,$ 
the collapse still can occur, albeit taking a longer actual time $t_{c}>T_{c}.$ 
(iii) {Increasing further, $\alpha$ reaches a critical value $\alpha_{\rm cr}\approx 0.7v_{c}$} such that 
at $\alpha>\alpha_{\rm cr}$ the anomalous velocity is large enough
to prevent the collapse. The dependence of $t_{c}$ on the SOC strength can be
described as $t_{c}\sim\left(\alpha_{\rm cr}-\alpha \right)^{-1}.$

To get an insight of the effects of SOC on
the collapse, we depict the density profiles in Fig. \ref{figure2}. At a large $\alpha$
the density forms a double peak with the maxima 
positions separating with time as a result of the centrifugal 
component in the flux. 
The resulting two-dimensional density distribution is given by a ring of  
radius $R(t)$ and width $w(t)$ with 
$a(t)\sim \sqrt{R(t)w(t)}$, responsible for the broad plateaus in $a(t)/a(0)$ 
seen in Fig. \ref{figure1} {at subcritical spin-orbit coupling.} 
At $R(t)\gg a(t)$ the interatomic interaction energy tends to zero as $-1/R(t)w(t)$, and the
conserved total energy is the sum of the kinetic and SOC
terms. At $\alpha<\alpha_{\rm cr},$ (see Fig. \ref{figure2}(a)) the attraction is still strong enough to
reverse the splitting and to restore the collapse. At $\alpha>\alpha_{\rm cr}$, (see Fig. \ref{figure2}(b)) the anomalous
velocity takes over, the splitting continues, and the collapse does not occur \cite{conf}. 
This process is naturally accompanied by evolution of the condensate flux and spin presented in the Supplemental
material \cite{supplement}.

\subsection{Balanced Rashba and Dresselhaus couplings} 

{In this subsection we consider a one-dimensional coupling}%
\begin{equation}
\wh{H}_{\mathrm{so}}\equiv \wh{H}_{RD}=\alpha k_{x}\widehat{\sigma }_{z},
\label{RDC}
\end{equation}%
which {is equivalent to} the balanced Rashba and Dresselhaus contributions and
gauged out by an $x-$dependent spin rotation \cite{Tokatly} ${\mathbf U}=\exp \left[ i%
\widehat{\sigma }_{z}x/L_{\mathrm{so}}\right] $.

For simplicity we consider the initial state corresponding to the spin oriented along the $x-$axis 
with ${\bm \psi}(0)=\left[1,1\right]^{T}/\sqrt{2}.$ 
Due to the anomalous velocity $\partial{\wh{H}_{RD}}/\partial k_{x}=\alpha\widehat{\sigma}_{z}$ 
(cf. Eq. \re{velocity}), the initial state splits into two
spin-projected wavepackets moving in the absence of interactions with velocities $\pm
\alpha.$ As a result, the effective interaction decreases, and the collapse can be
prohibited by this decrease. This happens, however, only at certain conditions, which
we establish here. For qualitative analysis we use the ansatz
\begin{equation}
\hspace{-.4cm}\psi ^{\uparrow, \downarrow }({\bf r}_{\mp},t)=\widetilde{A}(t)
\exp\left[-\frac{{r}_{\mp}^{2}}{2\widetilde{a}_{\rm v}^{2}(t)}
\left(1+i\widetilde{b}_{\rm v}(t)\right) \mp i\widetilde{c}_{\rm v}(t)x\right],
\end{equation}%
where the upper (lower) sign corresponds to spin up (down) and position ${\mathbf{r}}
_{\mp}\equiv (x\mp \widetilde{d}_{\rm v}(t),y)$, and in addition to the variational 
chirp $\widetilde{b}_{\rm v}(t)$ and width $\widetilde{a}_{\rm v}(t),$ we introduced 
the variational momentum $\widetilde{c}_{\rm v}(t)$. From this ansatz we obtain, 
using an approach similar to that of Ref. \cite{abdullaev},
equations of motion for $\widetilde{d}_{\rm v}$ and $\widetilde{a}_{\rm v}$:
\begin{eqnarray}
&&\overset{..}{\widetilde{d}}_{\rm v} = -\frac{\widetilde{g}_{2}N}{\pi }
\frac{\widetilde{d}_{\rm v}}{\widetilde{a}_{\rm v}^{4}}
\exp\left( -\frac{2\widetilde{d}_{\rm v}^{2}}{\widetilde{a}_{\rm v}^{2}}\right), \label{ansatzRD}\\
&&\overset{..}{\widetilde{a}}_{\rm v} = \frac{1}{\widetilde{a}_{\rm v}^{3}}\left[ \pi -\frac{\widetilde{g}_{2}N}{4}
\left[ 1+\left( 1-\frac{2\widetilde{d}_{\rm v}^{2}}{\widetilde{a}_{\rm v}^{2}}\right)
\exp\left( -\frac{2\widetilde{d}_{\rm v}^{2}}{\widetilde{a}_{\rm v}^{2}}\right) 
\right]\right], \nonumber
\end{eqnarray}%
for given ${\widetilde{a}}_{\rm v}(0)=a(0)$ and other initial conditions
\begin{equation}
{\widetilde{d}}_{\rm v}(0)=0,\quad\overset{.}{\widetilde{d}}_{\rm v}(0)=\alpha,\quad\overset{.}{\widetilde{a}}_{\rm v}(0)=0,
\end{equation}%
where $\overset{.}{\widetilde{d}}_{\rm v}(0)$ is due to the anomalous velocity term leading
to the spin-dependent splitting. 
These equations show that the collapse disappears if the coupling
is strong enough to sufficiently separate the spin components, that is 
at a certain time $\overset{..}{\widetilde{a}}_{\rm v}$ changes sign from negative to positive. 

Qualitative conditions of the collapse in the presence of spin-orbit
coupling in Eq. \re{RDC}, which can be found from Eq. \re{ansatzRD}, 
are as follows. If $\widetilde{g}_{2}N>4\pi$, 
the collapse always occurs since even if
the spin states are well-separated, each of them still has the sufficient number of
atoms. Depending on the interatomic interaction
and SOC, one can either obtain the collapse at the origin,
producing a spin non-polarized condensate, or two spatially 
symmetric ones producing $z-$axis polarized condensates. 
If $\widetilde{g}_{2}N<4\pi$, the collapse occurrence depends on the SOC strength.

At a sufficiently strong SOC, the spin splitting of the initial state  
and possible collapse happen on different time scales. The splitting occurs
fast, on the timescale of $a(0)/\alpha,$ and the interatomic attraction starts to play a role 
after the splitting. The condition of time scale separation, which
allows one to treat the splitting and the collapse independently, 
is formulated as $a(0)<T_{c}\alpha$ or, in other words, as
$\alpha >\sqrt{\Lambda }/a(0).$ This looks similar to the above condition
for the critical Rashba coupling. However, these conditions are
qualitatively different. For the Rashba coupling, the density decreases to
zero and the collapse disappears completely at any SOC
stronger than the critical one. For the balanced Rashba-Dresselhaus coupling
the maximum density decreases at most by a factor of two, and,
therefore, the collapse can occur even at a very strong SOC, when spin-up and spin-down
states are already well-separated in space.

\begin{figure}[t]
\begin{center}
\includegraphics[width=80mm]{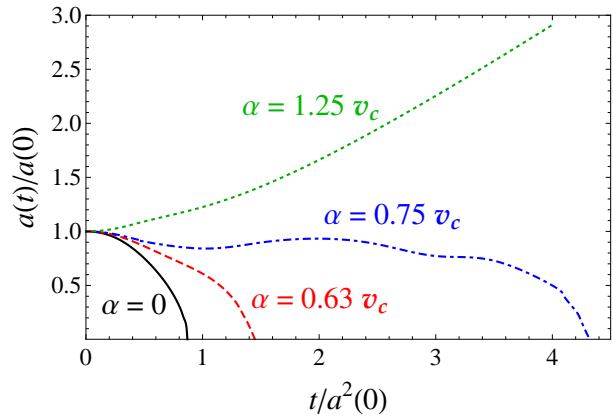}
\end{center}
\caption{(Color online) Time dependence of the condensate width  for $\widetilde{g}_{2}N=3\protect\pi$ 
in the presence of balanced Rashba-Dresselhaus SOC and different
values of $\alpha$ as marked near the lines.}
\label{figure3m}
\end{figure}

Figure \ref{figure3m} shows the time dependence of the packet width in Eq. \re{atime} obtained 
by solution of Eq. \re{SHE} with spin-orbit coupling Hamiltonian \re{RDC} for $\widetilde{g}_{2}N=3\pi.$
The behavior at small $t\ll T_{c}$ here depends on $\alpha$ since 
the peaks in the spin-projected densities split by $%
2\alpha\,t$ due to the anomalous velocity. The numerically obtained critical value of $\alpha$ here
is approximately $0.83 v_{c}$ and $a(t)\sim(t_{c}-t)$ shows a linear rather
than a square-root behavior near the collapse time.

\section{Relation to experiment and conclusions} 

To make connections to possible BEC experiments, we return to the physical units and 
estimate the constant $\widetilde{g}_{2}$
as $0.05$ for $-a_{s}\sim 100 a_{B}\sim 5\times10^{-3}$ $\mu{\rm m}$ and $a_{z}\sim 1$ $\mu{\rm m}$. The condition
$\widetilde{g}_{2}N>2\pi$ can be satisfied already for ensembles with $N\sim 100$ particles.
The velocity of the collapse is $v_{c}\sim \hbar\sqrt{\widetilde{g}_{2}N}/Ma(0).$ At $a(0)\sim 10$ $\mu{\rm m}$
and $N\sim10^{3}$ this estimate yields $v_{c}\sim 0.03$ cm/s and the corresponding time scale $T_{c}=a(0)/v_{c}\sim 0.3$ s.
Such a small value of $v_{c}$ demonstrates that even a relatively weak 
experimentally achievable coupling \cite{Campbell}  can prevent the BEC from collapsing. At these conditions, 
the characteristic distance between the particles $\left(a^{2}(0)a_{z}/N\right)^{1/3}\sim 0.5$ $\mu{\rm m}$ 
is much larger than $-4\pi a_{s}\le 0.1$ $\mu{\rm m},$ still preventing a strong depletion of the condensate. 
 
{To conclude, we have demonstrated that the anomalous spin-dependent velocity determined by the 
spin-orbit coupling  strength can prevent collapse of a nonuniform quasi two-dimensional BEC \cite{selftrap,polariton}.  
For the Rashba coupling with the spectrum axially symmetric in the momentum space, this velocity 
leads to a centrifugal component in the two-dimensional density flux. As a result, spin-orbit coupling can prevent
collapse of the two-dimensional BEC if this flux is sufficiently strong to overcome the effect of interatomic 
attraction. In this case, the attraction between the bosons cannot squeeze 
the initial wavepacket and force it to collapse. In the case of effectively one-dimensional balanced Rashba-Dresselhaus couplings,
the anomalous velocity splits the initial state into spin-polarized wave packets, decreases the condensate density and, thus, can prevent 
the collapse. Our approach can be generalized in a straightforward way for intermediate case, 
where Dresselhaus and Rashba couplings have different strength.}
These results show that one can 
gain a control over the BEC collapse process by using the experimentally available synthetic
spin-orbit coupling fields  {and, thus, extend the 
experimental abilities to study various nontrivial dynamical regimes in Bose-Einstein condensates of attracting particles.}

\section{Acknowledgement} 
This work was supported by the University of Basque Country UPV/EHU under
program UFI 11/55, Spanish MEC (FIS2012-36673-C03-01 and
FIS2012-36673-C03-03), and Grupos Consolidados UPV/EHU del Gobierno Vasco
(IT-472-10). S.M. acknowledges EU-funded Erasmus Mundus Action 2 eASTANA,
"evroAsian Starter for the Technical Academic Programme" (Agreement No.
2001-2571/001-001-EMA2). The research at Shanghai University 
was supported by the National Natural Science Foundation of China (11474193, 61404079 and 61176118), 
Shuguang Program (14SG35), the Pujiang, 
and Yangfan Program (13PJ1403000 and 14YF1408400), 
the Research Fund for the Doctoral Program (2013310811003), and the Program for Eastern Scholar.
We are grateful to M. Modugno and M. Glazov for valuable discussions and comments.

\newpage

\begin{widetext}

{\bf Supplemental Material for ``Collapse of spin-orbit coupled Bose-Einstein condensate''}


\section{Graphic material for strong interatomic interaction.}

Here we present additional graphic material to illustrate the role 
of spin-orbit coupling in the physics of the collapse of two-dimensional Bose-Einstein
condensates. 

\begin{figure}[h]
\begin{center}
\includegraphics[width=0.5\textwidth]{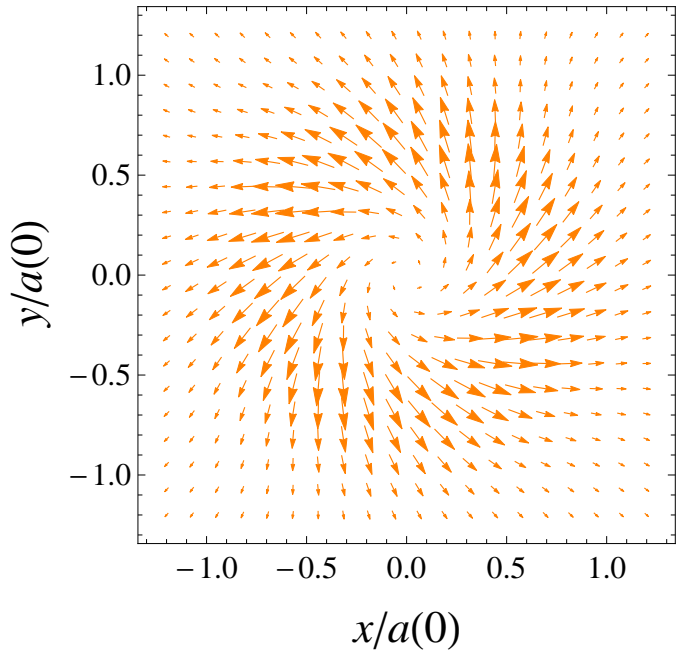}
\end{center}
\caption{Flux of the condensate, defined in Eq. (8) of the main text, at $\widetilde{g}_{2}N=16\pi$,  
$\alpha=0.84 v_{c}$, and $t=0.2a^{2}(0)$.
Since here $\alpha>\alpha_{\rm cr}=0.7v_{c},$ this is the no-collapse regime, corresponding 
to a plot in Fig. 2(b) of the main text.
This flux distribution leads to the increase in the width $a(t)$ as defined in Eq. (13) of the main text.}
\label{Figflux}
\end{figure}

\begin{figure}[h]
\begin{center}
\includegraphics[width=0.5\textwidth]{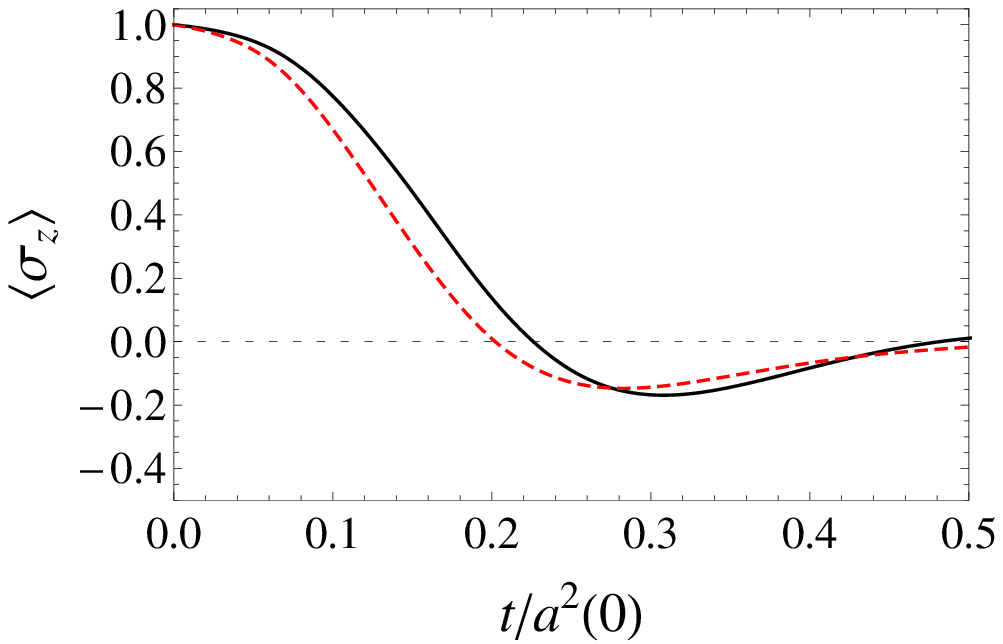}
\end{center}
\caption{ Time dependence of the total condensate spin component as defined by Eq. (9) in the main text. 
Here $\widetilde{g}_{2}N=16\pi$,
black solid line is for $\alpha=0.67 v_{c}<\alpha_{\rm cr}$ (collapse regime), and red dashed line is for $\alpha=0.84 v_{c}$ (no-collapse regime). 
}
\label{FigSigmaZStrong}
\end{figure}

\vspace{6cm}

\section{Near-the-threshold collapse}

Now we address a near-the-threshold collapse, which takes a long time, and where the difference between the 
variational $\widetilde{g}_{2}N=2\pi$ and the exact $\equiv\widetilde{g}_{2}N=1.862\pi$ threshold couplings becomes important.  
We take $\widetilde{g}_{2}N=2\pi,$ where in the absence of spin-related effects, the total energy in Eq. (3) of the main text
of a Gaussian state is zero. 
In this case the critical spin-orbit coupling $\alpha_{\rm cr}$ 
sufficient to destroy the collapse by the anomalous 
velocity is determined by condition $\alpha_{\rm cr}^{2}\sim v_{c}/a(0)$ and  
can be estimated as $\left(\widetilde{g}_{2}N-\lambda_{\rm ex}\right)^{1/4}.$ 
The numerically obtained critical value is $\alpha _{\rm cr}\approx 0.38(2\pi-\lambda_{\rm ex})^{1/4}$.
Figure \ref{figure3} shows that for $\alpha$ slightly larger than the critical value,
the condensate first narrows and then broadens. The peak structure of Fig. 2 in the main text  
is not formed here, and the collapse disappears due to the broadening rather than due to the splitting.
Although the change in the total spin shown in Fig. \ref{FigSigmaZ} is moderate 
compared to that presented in Fig. \ref{FigSigmaZStrong}, as expected for relatively small values of $\alpha,$ the 
collapse does not occur here.

\begin{figure}[h]
\begin{center}
\includegraphics[width=0.5\textwidth]{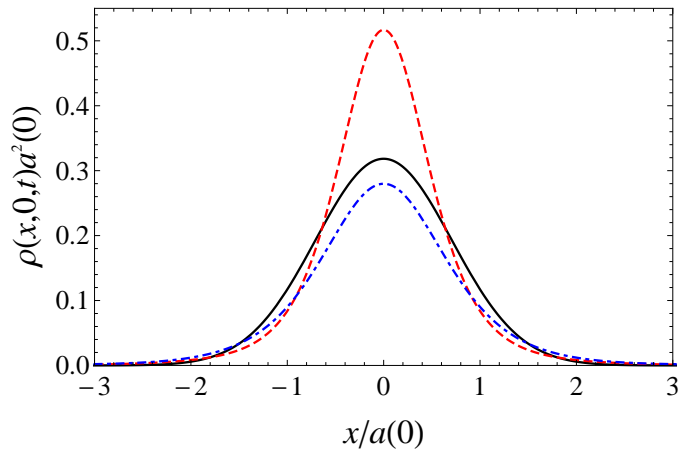}
\end{center}
\caption{(Color online) Density profile of the condensate for different times, $\widetilde{g}_{2}N=2\pi$;
$\protect\alpha=0.39(2\pi-\lambda_{\rm ex})^{1/4}$ (slightly above the critical value), black solid line is for $t=0$,
red dashed line is for $t=a^{2}(0),$ and blue dot-dashed line is for $t=4.4a^{2}(0)$. Here no collapse occurs.}
\label{figure3}
\end{figure}


\begin{figure}[h]
\begin{center}
\includegraphics[width=0.5\textwidth]{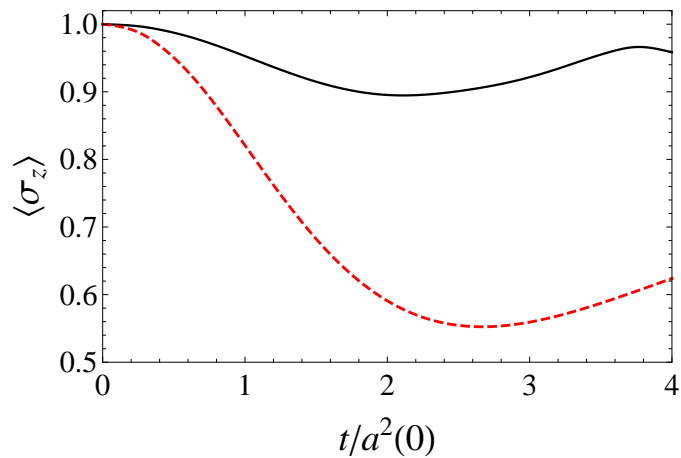}
\end{center}
\caption{ Time dependence of the total condensate spin component as defined by Eq. (9) in the main text. 
Here $\widetilde{g}_{2}N=2\pi$,
black solid line is for $\alpha=0.19(2\pi-\lambda_{\rm ex})^{1/4}$ (collapse regime) and 
red dashed line is for $\alpha=0.39(2\pi-\lambda_{\rm ex})^{1/4}$ (no-collapse regime).}
\label{FigSigmaZ}
\end{figure}

\end{widetext}



\begin{thebibliography}{99}
\bibitem{conf} The density can decrease to zero with infinite $R(t\rightarrow\infty)$ only 
in the absence of a strong confinement. {If the confinement potential 
$\omega_{0}^{2}\left(x^{2}+y^{2}\right)/2,$ where $\omega_{0}$ is the corresponding frequency,
is taken into account, the energy conservation limits the value of $R(t)$ and naturally increases 
the critical value of SOC.}
However, {as our simulations show,} in the case of a strong spin-orbit coupling, 
even if the density returns to the vicinity of the origin, where the confinement potential is weak,
the centrifugal flux takes over and 
the condensate begins to spread again rather than to collapse.
\bibitem{selftrap} In terms of collective effects in solids \cite{rashba_review}, this result implies that spin-orbit
coupling can prevent self-trapping of carriers and excitons.
\end{thebibliography}
\end{document}